\documentclass[twocolumn,prc,aps,nofootinbib]{revtex4-1}
\usepackage{amsmath,amssymb,slashed}
\usepackage[utf8]{inputenc}
\usepackage{hyperref}
\usepackage{natbib}
\usepackage{xcolor}
\usepackage[normalem]{ulem}

\newcommand{\mean}[1]{\langle #1 \rangle}

\newcommand{\D}{{\rm d}}
\newcommand{\I}{{\rm i}}
\def\spt{{\cal S}}
\def\wT{{\widehat T}}
\def\wj{{\widehat j}}
\def\wQ{{\widehat Q}}

\def\wP{{\widehat P}}
\def\wJ{{\widehat J}}

\def\wspt{{\widehat{\cal S}}}

\def\wPhi{{\widehat{\Phi}}}

\def\wrho{{\widehat{\rho}}}

\def\codev{{\stackrel{\leftrightarrow}{\partial}}}

\def\codevmu{{\stackrel{\leftrightarrow}{\partial}}\!\,^\mu}
\def\codevnu{{\stackrel{\leftrightarrow}{\partial}}\!\,^\nu}
\def\codevlambda{{\stackrel{\leftrightarrow}{\partial}}\!\,^\lambda}

\newcommand{\tr}{{\rm tr}}

\newcommand{\E}{{\rm e}}

\newcommand{\p}{{\rm p}}

\newcommand{\de}{\partial}
\newcommand{\subs}[1]{_{\textup{#1}}}
\newcommand{\sups}[1]{^{\textup{#1}}}
\newcommand{\be}{\begin{equation}}
	\newcommand{\ee}{\end{equation}}                                                                               
\newcommand{\bea}{\begin{eqnarray}}
	\newcommand{\eea}{\end{eqnarray}}

\newcommand{\h}[1]{\widehat{#1}}
\renewcommand{\vec}[1]{\ensuremath{\mathchoice				
		{\mbox{\boldmath$\displaystyle\mathbf{#1}$}}
		{\mbox{\boldmath$\textstyle\mathbf{#1}$}}
		{\mbox{\boldmath$\scriptstyle\mathbf{#1}$}}
		{\mbox{\boldmath$\scriptscriptstyle\mathbf{#1}$}}}}

\begin{document}
	
\title{Pseudogauge dependence of the spin polarization and of the axial vortical effect}
	
\author{M. Buzzegoli}\email{mbuzz@iastate.edu}\affiliation{Department of Physics and Astronomy, Iowa State University, Ames, Iowa 50011, USA}

\begin{abstract}
The mean spin polarization vector of spin 1/2 particles in a relativistic fluid at local thermal equilibrium (LTE) in different
pseudogauges (PGs), i.e., with different choices for the decomposition of orbital and spin angular momentum, is obtained.
The spin polarization obtained in the canonical PG differs from the one obtained in the Belinfante PG. It is found that this
difference can not be written by replacing the thermal vorticity with the spin potential in the usual polarization formula.
Other PG choices affect the contribution of thermal shear to the spin polarization. Therefore, the choice of a PG affects
the predictions for the polarization measured in heavy-ion collisions. In general, it is shown that the Wigner function of
a noninteracting Dirac field at LTE is affected by the PG transformations. Explicit expressions of the mean axial current
are also calculated and it is found that even the axial vortical effect conductivity depends on the PG.
\end{abstract}

\maketitle
\section{Introduction}
\label{intro}
The measurement of spin polarization of particles in the quark gluon plasma (QGP) formed in relativistic heavy ion collisions~\cite{star,Adam:2018ivw}
has opened the possibility for new phenomenological investigations of spin physics in relativistic fluids. Predictions of
relativistic hydrodynamics are able to explain the measurements of global (integrated over all momenta) spin polarization,
but fails to reproduce the data on momentum-dependent polarization~\cite{Adam:2019srw}; see~\cite{Becattini:2020ngo} for a review.
Recently, it was found that a previously overlooked spin-thermal shear coupling~\cite{Becattini:2021suc,Liu:2021uhn} could
restore the agreement between theory and experimental data~\cite{Becattini:2021iol,Fu:2021pok}. However, from the theory side,
one should take into account that the choice of a decomposition of the angular momentum into an orbital and spin part could significantly affect the
spin polarization predictions of a fluid at local thermal equilibrium.

For this reason, recent studies have begun to examine the role of the spin
tensor~\cite{Florkowski:2017ruc,Florkowski:2017dyn,Florkowski:2018ahw,Hattori:2019lfp,Bhadury:2020puc,Weickgenannt:2020aaf,Tinti:2020gyh,Fukushima:2020ucl,Hongo:2021ona,Bhadury:2021oat}
and also the impact of a spin potential~\cite{Florkowski:2018fap,Wu:2019eyi,Florkowski:2019gio,Gallegos:2020otk,Gallegos:2021bzp}
in relativistic hydrodynamics. In more general terms, different decompositions of the boost-angular momentum current are obtained
by choosing a particular form of the quantum energy-momentum tensor (EMT) $\wT^{\mu\nu}$ and of the spin tensor $\wspt^{\lambda,\mu\nu}$.
Indeed this choice is not unique in \emph{special} relativity. Given one pair of operators, one can always generate another equally
valid pair by means of a pseudogauge transformation (PGT)~\cite{Hehl:1976vr}:
\begin{equation*}
\begin{split}
	\wT^{\prime\mu\nu}= &\wT^{\mu\nu}+\frac{1}{2}\nabla_\lambda\left(\wPhi^{\lambda,\mu\nu}-\wPhi^{\mu,\lambda\nu} -\wPhi^{\nu,\lambda\mu}\right)\\
	\wspt^{\prime\lambda,\mu\nu}=&\wspt^{\lambda,\mu\nu}-\wPhi^{\lambda,\mu\nu}+ \nabla_\rho\h{Z}^{\mu\nu,\lambda\rho}\\
\end{split}
\end{equation*}
where the rank-3 tensor field $\wPhi$ and the rank-4 tensor $\h{Z}$ must satisfy the following symmetries:
\begin{equation*}
\wPhi^{\lambda,\mu\nu}=-\wPhi^{\lambda,\nu\mu},\quad \h{Z}^{\mu\nu,\lambda\rho}=-\h{Z}^{\nu\mu,\lambda\rho}=-\h{Z}^{\mu\nu,\rho\lambda}.
\end{equation*}
Since such transformations do not affect the equation of motions and the total energy, momentum, and angular momentum, they
are generally regarded as physically irrelevant. However, the relativistic hydrodynamic equations are obtained by
constraining local - as opposed to global - quantities, therefore the PGTs would lead to different physical
results. The role of PGTs in statistical quantum field theory was analyzed
here~\cite{Becattini:2011ev,Becattini:2012pp,Becattini:2018duy,Speranza:2020ilk}, where it was shown that they
might bring a contribution only when the system is out of global thermal equilibrium; see also~\cite{Li:2020eon}.

Despite the fact that this issue was already discussed in the literature, explicit formulas for spin polarization in different
pseudogauges have not yet been reported. Indeed, predictions for spin polarization of lambda hyperons produced in heavy-ion collisions are
obtained starting from the local thermal equilibrium statistical operator,
\begin{equation}
\label{eq:rhoLTEBintro}
	\wrho\subs{LTE}\sups{ B}=\frac{1}{\mathcal{Z}}\exp\left[-\int\D\Sigma_\mu\left( \wT^{\mu\nu}\subs{B}\beta_\nu-\wj^\mu\zeta\right)\right],
\end{equation}
obtained by choosing the Belinfante form of EMT $\wT\subs{B}$ with a corresponding vanishing spin tensor and where $\beta$ is
the fluid velocity $u$ divided by the temperature $T$.
The resulting spin polarization of a particle with momentum $k$ and mass $m$ at first order of thermodynamic
gradients is~\cite{Becattini:2013fla,Becattini:2020sww,Becattini:2021suc}
\begin{equation}
\label{eq:PolBelIntro}
S^\mu\subs{B}(k)\simeq S^\mu_{\varpi}(k) + S^\mu_{\xi}(k),
\end{equation}
where
\begin{equation}
\label{eq:PolVortIntro}
S^\mu_\varpi(k)= - \frac{1}{8m} \epsilon^{\mu\rho\sigma\tau} k_\tau \frac{\int_{\Sigma} \D \Sigma \cdot k \, n\subs{F} \left(1 -n\subs{F}\right) \varpi_{\rho\sigma}}
  {\int_{\Sigma} \D \Sigma \cdot k \, n\subs{F}},
\end{equation}
\begin{equation}
\label{eq:PolShearIntro}
S_\xi^\mu(k)=-\frac{1}{4m}\epsilon^{\mu\lambda\sigma\tau}\frac{ k_\tau k^\rho }{\varepsilon_k}
	 \frac{\int_\Sigma \D\Sigma\cdot k\, n\subs{F}\left(1-n\subs{F}\right)\hat{t}_\lambda\xi_{\rho\sigma}}{\int_\Sigma \D\Sigma \cdot k \, n\subs{F}},
\end{equation}
with $\varepsilon_k=\sqrt{\vec{k}^2+m^2}$, and
\begin{equation*}
\begin{split}
\varpi_{\mu\nu}=&-\frac{1}{2}\left( \de_\mu\beta_\nu- \de_\nu\beta_\mu\right),\\
\xi_{\mu\nu}=&\frac{1}{2}\left( \de_\mu\beta_\nu + \de_\nu\beta_\mu\right),
\end{split}
\end{equation*}
are respectively the thermal vorticity and the thermal shear, and $n\subs{F}=n\subs{F}(\beta(x)\cdot k)$
denotes the Fermi distribution function:
\begin{equation*}
n\subs{F}(z)=\frac{1}{\E^{z}+1}.
\end{equation*}
The QGP being a fluid out of equilibrium, one then expects PGTs to affect the prediction~\eqref{eq:PolBelIntro}.
It is then of phenomenological importance to quantify possible modifications to the previous formula in different pseudogauges.
In particular, the statistical operator obtained by choosing the canonical form of EMT $\wT\subs{C}$ and the associated canonical spin
tensor $\wspt\subs{C}$ is
\begin{equation}
\label{eq:rhoLTECanIntro}
\wrho\subs{LTE}\sups{C}=\frac{1}{\mathcal{Z}}\exp\left[-\int\D\Sigma_\mu\!\left( \wT^{\mu\nu}\subs{C}\beta_\nu
	\!-\frac{1}{2}\Omega_{\lambda\nu}\wspt\subs{C}^{\mu,\lambda\nu}\right)\right],
\end{equation}
where $\Omega$ is the spin potential. In this work I am going to show that if one starts from the statistical
operator~\eqref{eq:rhoLTECanIntro}, then the spin polarization becomes
\begin{equation}
\label{eq:PolCanIntro}
S^\mu\subs{C}(k)\simeq S^\mu\subs{B}(k) + \Delta\sups{C}_\Theta S^\mu(k),
\end{equation}
which differs from~\eqref{eq:PolBelIntro} by
\begin{equation}
\label{eq:DeltaSCanIntro}
\begin{split}
\Delta\sups{C}_\Theta S^\mu(k)=& \frac{\epsilon^{\lambda\rho\sigma\tau}\hat{t}_\lambda(k^\mu k_\tau-\eta^\mu_{\;\tau}m^2)}{8 m \varepsilon_k}\\
	&\times \frac{\int_\Sigma \D\Sigma(x)\cdot k\, n\subs{F}\left(1-n\subs{F}\right)\left(\varpi_{\rho\sigma}-\Omega_{\rho\sigma}\right)}{\int_\Sigma \D\Sigma \cdot k \, n\subs{F}},
\end{split}
\end{equation}
where $\hat{t}$ is the time direction in the fluid frame. I will also show that either the
de Groot-van Leeuwen-van Weert and Hilgevoord-Wouthuysen (HW) decompositions of angular momentum
result in the spin polarization
\begin{equation*}
\begin{split}
S^\mu\subs{GLW,HW}(k)\simeq & S^\mu\subs{C}(k) + \Delta\sups{GLW,HW}_\Theta S^\mu(k) +  \Delta\sups{GLW,HW}_\xi S^\mu(k)\\
=&- \frac{1}{8m} \epsilon^{\mu\rho\sigma\tau} k_\tau \frac{\int_{\Sigma} \D \Sigma \cdot k \, n\subs{F} \left(1 -n\subs{F}\right) \Omega_{\rho\sigma}}
  {\int_{\Sigma} \D \Sigma \cdot k \, n\subs{F}}
\end{split}
\end{equation*}
with
\begin{equation}
\label{eq:DeltaThetaGLWIntro}
\begin{split}
\Delta\sups{GLW,HW}_\Theta S^\mu(k) = & -\frac{1}{4m}\epsilon^{\mu\lambda\rho\tau}\hat{t}_\lambda\frac{ k_\tau k^\sigma }{\varepsilon_k}\\
	&\times \frac{\int_\Sigma \D\Sigma\cdot k\, n\subs{F}\left(1-n\subs{F}\right) \left(\varpi_{\rho\sigma}-\Omega_{\rho\sigma}\right)}{\int_\Sigma \D\Sigma \cdot k \, n\subs{F}},
\end{split}
\end{equation}
\begin{equation}
\label{eq:DeltaXiGLWIntro}
\begin{split}
\Delta\sups{GLW,HW}_\xi S^\mu(k) = &-S_\xi^\mu(k)=+\frac{1}{4m}\epsilon^{\mu\lambda\sigma\tau}\frac{ k_\tau k^\rho }{\varepsilon_k}\\
	&\times \frac{\int_\Sigma \D\Sigma\cdot k\, n\subs{F}\left(1-n\subs{F}\right)\hat{t}_\lambda\xi_{\rho\sigma}}{\int_\Sigma \D\Sigma \cdot k \, n\subs{F}}
\end{split}
\end{equation}
and where the expression was simplified by taking advantage of the Schouten identity.
Therefore the predictions of spin polarization in heavy-ion collisions depend on the chosen PG. Notice that
modifications~\eqref{eq:DeltaSCanIntro} and \eqref{eq:DeltaThetaGLWIntro} rely on the spin potential and vanish
when $\Omega=\varpi$, while modifications such as~\eqref{eq:DeltaXiGLWIntro} do not require the presence of a spin potential.
Similarly, I will show that the axial vortical effect (AVE)~\cite{Vilenkin:1979ui,Landsteiner:2011iq,Gao:2012ix,Kharzeev:2015znc},
which is the mean axial current of a fermion induced along the rotation of the fluid, is also affected by the PGT in
a system out of equilibrium.

The paper is organized as follows: in Sec.~\ref{sec:LocalEquilibrium} I review how to obtain the statistical density operator
in Zubarev's statistical quantum field theory and how it is affected by PGTs. In Sec.~\ref{sec:Wigner}
I consider a free Dirac field and, by using linear response theory, I evaluate the differences between the Wigner function
resulting from the Belinfante decomposition of angular momentum and the Wigner functions resulting from other pseudogauges.
In Sec.~\ref{sec:Polarization} I obtain the different predictions of spin polarization in various pseudogauges and I discuss
their relevance in heavy-ion collisions. In Sec.~\ref{sec:AxialC} I evaluate the mean axial current of a free Dirac field
and I show that the axial vortical effect conductivity also depend on the PGTs. Finally, in Sec.~\ref{sec:conclusion}
I summarize and discuss these findings.

\subsection*{Notation}
In this paper I adopt the natural units, with $\hbar=c=k_B=1$. The Minkowskian metric tensor $g$ is ${\rm diag}(1,-1,-1,-1)$;
for the Levi-Civita symbol I use the convention $\epsilon^{0123}=1$.\\ 
I use the relativistic notation with repeated indices assumed to be saturated. Operators in Hilbert 
space are denoted by a wide upper hat, e.g. $\widehat H$, except the Dirac field operator which is
denoted by a $\Psi$. The names ``mean value'', ``(thermal) expectation value'' and ``(thermal) average''
of an operator $\h{O}$ are all used interchangeably to denote the trace with a statistical
operator $\wrho$: $\tr\left(\wrho\,\h{O}\right)$.

\section{Local thermal equilibrium statistical operators}
\label{sec:LocalEquilibrium}
The statistical operator of a relativistic quantum fluid that has reached the local thermodynamic equilibrium (LTE)
can be obtained by Zubarev’s method~\cite{Zubarev:1979,vanWeert}. This approach is particularly suitable for
the quark gluon plasma produced in heavy-ion collisions and, among other predictions, it has been used to obtain the spin polarization
formula~\eqref{eq:PolBelIntro}~\cite{Becattini:2020sww}. A more detailed review of the method can be found in~\cite{Becattini:2019dxo,Buzzegoli:2020ycf}.
Here, I want to highlight the effect of pseudogauge transformations on the form of the statistical operator, as already discussed
in~\cite{Becattini:2018duy,Speranza:2020ilk}.

If the system reached LTE at a certain time $\tau_0$, then one can describe it with an energy-momentum density $T^{\mu\nu}$,
a boost-angular momentum density $\mathcal{J}^{\mu,\lambda\nu}$, and an (electric) conserved current $j^\mu$, all lying on a
space-like hyper-surface $\Sigma(\tau_0)$. The LTE density operator $\wrho$ is then obtained by maximizing the entropy
$S=-\tr(\wrho \log \wrho)$ with the constraints of the mean conserved currents being equal to the actual ones in the
hyper-surface $\Sigma(\tau_0)$. These constraints are obtained by projecting the mean values of the quantum operators
onto $n$, the normalized vector perpendicular to $\Sigma(\tau_0)$:
\begin{equation}
\label{eq:Constr}
\begin{split}
n_\mu& \tr\left(\wrho\, \wT^{\mu\nu} \right)=n_\mu T^{\mu\nu},\\
n_\mu& \tr\left(\wrho\, \h{\mathcal{J}}^{\mu,\lambda\nu} \right)=n_\mu \mathcal{J}^{\mu,\lambda\nu},\\
n_\mu& \tr\left(\wrho\, \wj^\mu \right)=n_\mu j^\mu,
\end{split}
\end{equation}
where the operators are in the Heisenberg representation. Notice that since the
boost-angular momentum current is written in terms of the EMT and of the spin tensor as
\begin{equation*}
\h{\mathcal{J}}^{\mu,\lambda\nu}=x^\lambda \wT^{\mu\nu} -x^\nu \wT^{\mu\lambda} + \wspt^{\mu,\lambda\nu},
\end{equation*}
the constraint on EMT is redundant and the boost-angular momentum is fixed just by constraining the spin tensor
\begin{equation*}
	n_\mu \tr\left(\wrho\, \wspt^{\lambda,\mu\nu} \right)=n_\lambda \spt^{\lambda,\mu\nu}.
\end{equation*}
Then, the resulting statistical operator is
\begin{equation*}
\wrho=\!\frac{1}{\mathcal{Z}}\exp\left[-\!\int_{\Sigma(\tau_0)}\!\!\!\!\!\!\!\!\D\Sigma_\mu\!\left( \wT^{\mu\nu}\beta_\nu
	\!-\frac{1}{2}\Omega_{\lambda\nu}\wspt^{\mu,\lambda\nu}\!-\!\wj^\mu\zeta\right)\right],
\end{equation*}
where $\beta$, $\Omega$, and $\zeta$ are the Lagrange multipliers of the problem and they have the physical meaning of
four-temperature vector, spin potential, and chemical potential divided by the temperature $T=1/\sqrt{\beta^2}$, respectively~\cite{Becattini:2014yxa}.
Notice that statistical operator is not time-independent and that it can be written by means of Gauss's theorem as~\cite{Becattini:2014yxa,Becattini:2020sww,Becattini:2021suc}
\begin{equation}
\label{eq:statoperTrue}
\begin{split}
\wrho=&\!\frac{1}{\mathcal{Z}}\exp\left[-\!\int_{\Sigma(\tau)}\!\!\!\!\!\!\!\!\D\Sigma_\mu\!\left( \wT^{\mu\nu}\beta_\nu
	\!-\frac{1}{2}\Omega_{\lambda\nu}\wspt^{\mu,\lambda\nu}\!-\!\wj^\mu\zeta\right)\right.\\
&+\left.\int_V\!\!\D V \nabla_\mu\left( \wT^{\mu\nu}\beta_\nu -\frac{1}{2}\Omega_{\lambda\nu}\wspt^{\mu,\lambda\nu}-\!\wj^\mu\zeta\right) \right],
\end{split}
\end{equation}
where $\Sigma(\tau)$ is the space-like hyper-surface at ``time'' $\tau$, and $V$ is the region of space-time
encompassed by the space-like hypersurfaces $\Sigma(\tau_0)$, $\Sigma(\tau)$ and the time-like boundaries.
The dissipative content of the system is only contained in the second term of the exponent~\cite{Becattini:2019dxo}.
For a quasi-ideal fluid such as the QGP, one expects the first term to be the predominant one, while the second term is supposedly
a correction. This second term requires a careful analysis and is left for future research. In the linear
response theory, it is always possible to include these neglected
dissipative effects later by adding them together with the non-dissipative ones.
The statistical operator is then only given by the first term in the
exponent, which is the local thermal equilibrium form at time $\tau$ and only contains non-dissipative effects:
\begin{equation}
\label{eq:statoperLTE}
\wrho\subs{LTE}=\!\frac{1}{\mathcal{Z}}\exp\left[-\!\int_{\Sigma(\tau)}\!\!\!\!\!\!\!\!\D\Sigma_\mu\left( \wT^{\mu\nu}\beta_\nu
	-\frac{1}{2}\Omega_{\lambda\nu}\wspt^{\mu,\lambda\nu}-\!\wj^\mu\zeta\right)\right].
\end{equation}

The relativistic nature of the fluid requires that the constraints~\eqref{eq:Constr} were given in terms of
the local currents. As long as the global equilibrium conditions are not enforced (see discussion below),
it is expected that the form of the quantum operators in~\eqref{eq:Constr} affects the form of the
statistical operator~\eqref{eq:statoperLTE}. Therefore, I now evaluate the effect of pseudogauge transformations
(PGTs) on the LTE statistical operator~\eqref{eq:statoperLTE}.

Suppose one repeats the procedure above with the symmetric Belinfante EMT $\wT^{\mu\nu}\subs{B}$. The spin tensor
associated with the Belinfante pseudogauge is a vanishing one ($\wspt^{\lambda,\mu\nu}\subs{B}=0$) and the total
boost-angular momentum current has the form of an orbital angular momentum:
\begin{equation*}
\mathcal{J}^{\mu,\lambda\nu}\subs{B}=x^\lambda \wT^{\mu\nu}\subs{B} -x^\nu \wT^{\mu\lambda}\subs{B}.
\end{equation*}
In this particular case, the constraint on the boost-angular momentum current is unnecessary as it is already being taken
care of in the constraint on the energy-momentum current. The resulting statistical operator is
\begin{equation}
\label{eq:rhoLTEB}
	\wrho\subs{LTE}\sups{ B}=\frac{1}{\mathcal{Z}}\exp\left[-\int\D\Sigma_\mu\left( \wT^{\mu\nu}\subs{B}\beta_\nu-\wj^\mu\zeta\right)\right].
\end{equation}

In a general pseudogauge the EMT is not symmetric and the spin tensor is not vanishing.
It is always possible to obtain any generic EMT and spin tensor $\wT^{\mu\nu}_\Phi$ and $\wspt_\Phi^{\lambda,\mu\nu}$
from the Belinfante one with the pseudogauge transformation:
\begin{equation}\label{eq:PsGaugeBel}
\begin{split}
	\wT^{\mu\nu}_\Phi= &\wT^{\mu\nu}\subs{B}+\frac{1}{2}\nabla_\lambda\left(\wPhi^{\lambda,\mu\nu}-\wPhi^{\mu,\lambda\nu} -\wPhi^{\nu,\lambda\mu}\right),\\
	\wspt^{\lambda,\mu\nu}_\Phi=&-\wPhi^{\lambda,\mu\nu}+ \nabla_\rho\h{Z}^{\mu\nu,\lambda\rho},\\
\end{split}
\end{equation}
for some specific $\wPhi$ and $\h{Z}$. Consider the LTE statistical operator~\eqref{eq:statoperLTE} obtained with
a specific choice of pseudogauge:
\begin{widetext}
\begin{equation}
\label{eq:genLETphi}
\wrho\subs{LTE}^{\,\Phi}=\frac{1}{\mathcal{Z}}\exp\left[-\int\D\Sigma_\mu\!\left( \wT^{\mu\nu}_\Phi\beta_\nu
	\!-\frac{1}{2}\Omega_{\lambda\nu}\wspt^{\mu,\lambda\nu}_\Phi\!-\wj^\mu\zeta\right)\right].
\end{equation}
The statistical operator~\eqref{eq:genLETphi} can be written in terms of the Belinfante EMT by taking advantage
of the transformations~\eqref{eq:PsGaugeBel}:
\begin{equation*}
\begin{split}
\wrho\subs{LTE}^{\,\Phi}= & \frac{1}{\mathcal{Z}}\exp\left\{-\int\D\Sigma_\mu\left[ \wT^{\mu\nu}\subs{B}\beta_\nu
	+\frac{1}{2}\nabla_\lambda\left(\wPhi^{\lambda,\mu\nu}-\wPhi^{\mu,\lambda\nu} -\wPhi^{\nu,\lambda\mu}\right)\beta_\nu
	+\frac{1}{2}\Omega_{\lambda\nu}\left(\wPhi^{\mu,\lambda\nu}-\nabla_\rho\h{Z}^{\lambda\nu,\mu\rho}\right)-\wj^\mu\zeta\right]\right\}\\
=&\frac{1}{\mathcal{Z}}\exp\left\{-\int\D\Sigma_\mu\left[ \wT^{\mu\nu}\subs{B}\beta_\nu
	+\frac{1}{2}\nabla_\lambda\left(\beta_\nu\wPhi^{\lambda,\mu\nu}-\beta_\nu\wPhi^{\mu,\lambda\nu} -\beta_\nu\wPhi^{\nu,\lambda\mu}\right)+\right.\right.\\
&-\left.\left.\frac{1}{2}\nabla_\lambda\beta_\nu \left(\wPhi^{\lambda,\mu\nu}-\wPhi^{\mu,\lambda\nu} -\wPhi^{\nu,\lambda\mu}\right)
	+\frac{1}{2}\Omega_{\lambda\nu}\wPhi^{\mu,\lambda\nu}-\frac{1}{2}\Omega_{\lambda\nu}\nabla_\rho\h{Z}^{\lambda\nu,\mu\rho}-\wj^\mu\zeta\right]\right\}.
\end{split}
\end{equation*}
\end{widetext}
The second term in the exponent is a total divergence and is vanishing for suitable boundary conditions imposed
on $\beta$ and/or $\wPhi$. Then, by splitting the $\beta$ derivative into a symmetric and anti-symmetric parts,
\begin{equation*}
\begin{split}
\varpi_{\mu\nu}=&-\frac{1}{2}\left( \nabla_\mu\beta_\nu- \nabla_\nu\beta_\mu\right),\\
\xi_{\mu\nu}=&\frac{1}{2}\left( \nabla_\mu\beta_\nu + \nabla_\nu\beta_\mu\right),
\end{split}
\end{equation*}
one obtains
\begin{equation}
\label{eq:rhoLTEphi}
\begin{split}
\wrho\subs{LTE}^{\,\Phi}=&\frac{1}{\mathcal{Z}}\exp\left\{\!-\!\int\!\!\D\Sigma_\mu\!\!\left[ \wT^{\mu\nu}\subs{B}\beta_\nu\!
	-\!\frac{1}{2}\left(\varpi_{\lambda\nu}-\Omega_{\lambda\nu}\right)\wPhi^{\mu,\lambda\nu}+\right.\right.\\
	&\left.\left.- \xi_{\lambda\nu} \wPhi^{\lambda,\mu\nu}-\frac{1}{2}\Omega_{\lambda\nu}\nabla_\rho\h{Z}^{\lambda\nu,\mu\rho}-\wj^\mu\zeta\right]\right\}.
\end{split}
\end{equation}
The anti-symmetric derivative of the $\beta$ field $\varpi$ is called the thermal vorticity and contains information
about the acceleration and the rotation of the fluid~\cite{Becattini:2015nva}. Instead, the symmetric derivative $\xi$
is called the thermal shear tensor, see~\cite{Becattini:2021suc,Liu:2021uhn}.
The relation between the statistical operators obtained in different pseudogauges was discussed in~\cite{Becattini:2018duy,Speranza:2020ilk}.

As a notable example, the canonical EMT and spin tensor are related to the Belinfante one with the PGT~\eqref{eq:PsGaugeBel} with
\begin{equation*}
	\wPhi^{\lambda,\mu\nu}\subs{C}=-\wspt\subs{C}^{\lambda,\mu\nu},\quad
	\h{Z}^{\mu\nu,\lambda\rho}\subs{C}=0,
\end{equation*}
where $\wspt\subs{C}$ is the canonical spin tensor which is obtained directly by the Noether theorem:
\begin{equation*}
\wspt\subs{C}^{\lambda,\mu\nu}= -\I \sum_{a,b} \frac{\delta\mathcal{L}}{\delta(\de_\lambda\h{\psi}^a)}D(J^{\mu\nu})^a_{\hphantom{a}b}\h{\psi}^b
\end{equation*}
with D being the irreducible representation matrix of the Lorentz group pertaining to the field. In this case,
the statistical operator becomes:
\begin{equation*}
\begin{split}
\wrho\subs{LTE}\sups{ C}=&\frac{1}{\mathcal{Z}}\exp\left\{\!-\!\int\!\!\D\Sigma_\mu\!\!\left[ \wT^{\mu\nu}\subs{B}\beta_\nu\!
	+\!\frac{1}{2}\left(\varpi_{\lambda\nu}-\Omega_{\lambda\nu}\right)\wspt\subs{C}^{\mu,\lambda\nu}+\right.\right.\\
	&\left.\left.+ \xi_{\lambda\nu} \wspt\subs{C}^{\lambda,\mu\nu}-\wj^\mu\zeta\right]\right\}.
\end{split}
\end{equation*}
The differences between this operator obtained in the canonical pseudogauge and the one obtained in the Belinfante one
are discussed in ref.~\cite{Becattini:2018duy}.

Besides the canonical decomposition, infinite choices for $\wPhi$ and $\h{Z}$ are possible.
For instance, if there exist a vector and/or an axial currents $\wj\subs{V}^\mu$
and $\wj\subs{A}^\mu$, other possible choices for $\wPhi$ are given by
\begin{equation*}
\begin{split}
	\wPhi^{\lambda,\mu\nu}_{\epsilon{\rm A}}=&\frac{1}{2}\epsilon^{\lambda\tau\mu\nu}\wj_{{\rm A}\,\tau},\quad
	\wPhi^{\lambda,\mu\nu}\subs{A}=\eta^{\lambda\mu}\wj^\nu\subs{A}-\eta^{\lambda\nu}\wj\subs{A}^\mu,\\
	\wPhi^{\lambda,\mu\nu}_{\epsilon{\rm V}}=&\frac{1}{2}\epsilon^{\lambda\tau\mu\nu\sigma}\wj_{{\rm V}\,\tau},\quad
	\wPhi^{\lambda,\mu\nu}\subs{V}=\eta^{\lambda\mu}\wj\subs{V}^\nu-\eta^{\lambda\nu}\wj\subs{V}^\mu,
\end{split}
\end{equation*}
or linear combination of all the above. Taking advantage of the derivative operator, one can also choose
\begin{equation*}
\begin{split}
	\wPhi^{\lambda,\mu\nu}_{\epsilon\de}=&\frac{\I}{2}\epsilon^{\lambda\tau\mu\nu}\h{\de}_\tau,\quad
	\wPhi^{\lambda,\mu\nu}_\de=\I\eta^{\lambda\mu}\h{\de}^\nu-\I\eta^{\lambda\nu}\h{\de}^\mu.
\end{split}
\end{equation*}

This work will focus on the noninteracting Dirac field $\Psi$.  In this case, I will
discuss the following choices for $\wPhi$:
\begin{equation}
\label{eq:PGPotentialsGamma}
\begin{split}
\wPhi^{\lambda,\mu\nu}\subs{C}=&\wPhi^{\lambda,\mu\nu}_{\epsilon{\rm A}}=\frac{1}{2}\epsilon^{\lambda\mu\nu\tau}\bar\Psi\gamma_\tau\gamma^5\Psi\\
		=&-\frac{\I}{8}\bar\Psi\left\{\gamma^\lambda,\left[\gamma^\mu,\gamma^\nu\right]\right\}\Psi,\\
\wPhi^{\lambda,\mu\nu}_{\epsilon{\rm V}}=&\frac{1}{2}\epsilon^{\lambda\mu\nu\tau}\bar\Psi \gamma_\tau\Psi,\\
\wPhi^{\lambda,\mu\nu}\subs{A}=&\bar\Psi\left(\eta^{\lambda\mu}\gamma^\nu\gamma_5-\eta^{\lambda\nu}\gamma^\mu\gamma_5\right)\Psi,\\
\wPhi^{\lambda,\mu\nu}\subs{V}=&\bar\Psi\left(\eta^{\lambda\mu}\gamma^\nu-\eta^{\lambda\nu}\gamma^\mu\right)\Psi,
\end{split}
\end{equation}
where $\gamma$ are the Dirac Gamma matrices. I will also consider
\begin{equation}
\label{eq:PGPotentialsDe}
\begin{split}
	\wPhi^{\lambda,\mu\nu}_{\epsilon\de}=&\frac{\I}{2}\epsilon^{\lambda\tau\mu\nu}\bar\Psi \codev_\tau \Psi,\\
	\wPhi^{\lambda,\mu\nu}_\de=&\I\bar\Psi\left(\eta^{\lambda\mu}\codevnu-\eta^{\lambda\nu}\codevmu\right)\Psi,\\
	\wPhi^{\lambda,\mu\nu}_{\epsilon\de {\rm A}}=&\frac{1}{2}\epsilon^{\lambda\tau\mu\nu}\bar\Psi \codev_\tau \gamma_5\Psi,\\
	\wPhi^{\lambda,\mu\nu}_{\de {\rm A}}=&\bar\Psi\left(\eta^{\lambda\mu}\codevnu-\eta^{\lambda\nu}\codevmu\right)\gamma_5\Psi,\\
	\wPhi^{\lambda,\mu\nu}_{\de\Sigma}=& \frac{\I}{m}\bar\Psi\codevlambda\sigma^{\mu\nu}\Psi,
\end{split}
\end{equation}
where
\begin{equation*}
\codevmu=\left({\stackrel{\rightarrow}{\partial}}\!\,^\mu-{\stackrel{\leftarrow}{\partial}}\!\,^\mu\right),\quad
\sigma^{\mu\nu}=\frac{\I}{2}\left[\gamma^\mu,\gamma^\nu\right].
\end{equation*}
Other than the canonical and Belinfante pseudogauges, other valid choices of boost-angular momentum decomposition
for the Dirac field are given by the de Groot-van Leeuwen-van Weert (GLW) and the Hilgevoord-Wouthuysen (HW) decomposition.
The de Groot-van Leeuwen-van Weert EMT and spin tensor (see~\cite{Groot}) are obtained from the Belinfante ones
with~\cite{Speranza:2020ilk}
\begin{equation}
\label{eq:PGPotentialGLW}
\begin{split}
\wPhi^{\lambda,\mu\nu}\subs{GLW}=&-\wspt\subs{C}^{\lambda,\mu\nu}+\frac{\I}{4m}\bar\Psi\left(\sigma^{\lambda\mu}\codevnu-\sigma^{\lambda\nu}\codevmu\right)\Psi,\\
\quad \h{Z}^{\mu\nu,\lambda\rho}\subs{GLW}=&0.
\end{split}
\end{equation}
Instead, the Hilgevoord-Wouthuysen EMT and spin tensor are obtained with~\cite{Hehl:1997ep,Kirsch:2001gt,Speranza:2020ilk}
\begin{equation}
\label{eq:PGPotentialHW}
\begin{split}
\wPhi^{\lambda,\mu\nu}\subs{HW}=&\wPhi^{\lambda,\mu\nu}\subs{GLW}
	-\frac{\I}{4m}\bar\Psi\left(\eta^{\lambda\mu}\sigma^{\nu\alpha}-\eta^{\lambda\nu}\sigma^{\mu\alpha}\right)\codev_\alpha\Psi,\\
\h{Z}^{\mu\nu,\lambda\rho}\subs{HW}=&-\frac{1}{8m}\bar\Psi\left(\sigma^{\mu\nu}\sigma^{\lambda\rho}+\sigma^{\lambda\rho}\sigma^{\mu\nu}\right)\Psi.
\end{split}
\end{equation}
Lastly, I will also consider the improved EMT proposed by Callan, Coleman, and Jackiew (CCJ)~\cite{CCJ:1970}, whose matrix elements are
finite at all orders of perturbation theory for a renormalizable interaction. The improved EMT is obtained with
\begin{equation}
\label{eq:PGPotentialCCJ}
\begin{split}
\wPhi^{\lambda,\mu\nu}\subs{CCJ}=&-\frac{1}{6}\left(\eta^{\lambda\nu}\partial^\mu-\eta^{\lambda\mu}\partial^\nu\right)\bar\Psi\Psi,\\
\h{Z}^{\mu\nu,\lambda\rho}\subs{CCJ}=&-\frac{1}{6}\left(\eta^{\rho\mu}\eta^{\lambda\nu}-\eta^{\rho\nu}\eta^{\lambda\mu}\right)\bar\Psi\Psi,
\end{split}
\end{equation}
so that
\begin{equation*}
\begin{split}
\wspt^{\lambda,\mu\nu}\subs{CCJ}=&0,\\
\wT^{\mu\nu}\subs{CCJ}=&\wT^{\mu\nu}\subs{B}-\frac{1}{6}\left(\de^\mu\de^\nu-\eta^{\mu\nu}\de_\lambda\de^\lambda\right)\bar\Psi\Psi.
\end{split}
\end{equation*}
%

\subsection{Global equilibrium}
\label{subsec:global}
If the system is at the global thermal equilibrium, the statistical operator only depends on the global conserved quantities,
i.e. the total momentum, boost-angular momentum and the conserved charge.  Since these quantities are not
affected by the PGTs, the statistical operator of global equilibrium is also pseudogauge independent~\cite{Becattini:2018duy,Speranza:2020ilk}.
In what follows, I review how this comes to be starting from the local thermodynamic equilibrium statistical operator~\eqref{eq:rhoLTEphi}.

To describe a global thermal equilibrium one needs to impose that the statistical operator is actually time-independent, which occurs when the
integrals in~\eqref{eq:rhoLTEphi} are independent of the space-like hypersurfaces $\Sigma$. This is equivalent to imposing that
the divergence of the integrand is vanishing~\cite{Becattini:2012tc}:
\begin{equation*}
\begin{split}
\nabla_\mu& \left[ \wT^{\mu\nu}\subs{B}\beta_\nu\!
	-\!\frac{1}{2}\left(\varpi_{\lambda\nu}-\Omega_{\lambda\nu}\right)\wPhi^{\mu,\lambda\nu}+\right.\\
	&\left.- \xi_{\lambda\nu} \wPhi^{\lambda,\mu\nu}-\frac{1}{2}\Omega_{\lambda\nu}\nabla_\rho\h{Z}^{\lambda\nu,\mu\rho}-\wj^\mu\zeta\right]=0.
\end{split}
\end{equation*}
Taking advantage of the operatorial equations,
\begin{equation*}
\begin{split}
	\nabla_\mu \wT^{\mu\nu}\subs{B}\beta_\nu =& 0,\quad \nabla_\mu \wj^\mu=0,\\
	\nabla_\mu \wPhi^{\mu,\lambda\nu}=& \wT^{\lambda\nu}_\Phi - \wT^{\nu\lambda}_\Phi,\quad
	\nabla_\mu\nabla_\rho\h{Z}^{\lambda\nu,\mu\rho}=0,
\end{split}
\end{equation*}
the divergence of the integrand is written as
\begin{equation*}
\begin{split}
&\wT^{\mu\nu}\subs{B}\nabla_\mu\beta_\nu -\frac{1}{2}\left(\varpi_{\lambda\nu}-\Omega_{\lambda\nu}\right)\left(\wT^{\lambda\nu}_\Phi - \wT^{\nu\lambda}_\Phi\right)+\\
&-\frac{1}{2}\wPhi^{\mu,\lambda\nu}\nabla_\mu \left(\varpi_{\lambda\nu}-\Omega_{\lambda\nu}\right)
	-\frac{1}{2}\left(\nabla_\rho\h{Z}^{\lambda\nu,\mu\rho}\right)\nabla_\mu \Omega_{\lambda\nu}+\\
&- \xi_{\lambda\nu} \nabla_\mu\wPhi^{\lambda,\mu\nu} -  \wPhi^{\lambda,\mu\nu}\nabla_\mu\xi_{\lambda\nu}-\wj^\mu\nabla_\mu\zeta=0.
\end{split}
\end{equation*}
The last term of the divergence is vanishing if $\nabla_\mu \zeta=0$. In flat space-time this condition sets $\zeta$ to be a constant.
The first term is vanishing if $\beta$ is a Killing field, i.e., if it solves
\begin{equation*}
\nabla_\mu\beta_\nu+\nabla_\nu\beta_\mu=0.
\end{equation*}
In flat space-time the most general solution of the Killing equation is given by
\begin{equation*}
	\beta^\mu=b_\mu + \varpi_{\mu\nu}x^\nu
\end{equation*}
with $b$ a constant four-vector and $\varpi$ a constant anti-symmetric rank-2 tensor. This global equilibrium form of $\beta$
implies $\xi_{\lambda\nu}=0$ and $\nabla_\mu\xi_{\lambda\nu}=0$. Then, the integrand is vanishing under the following conditions:
\begin{equation*}
\begin{cases}
	\nabla_\mu \left(\varpi_{\lambda\nu}-\Omega_{\lambda\nu}\right)=\nabla_\mu \Omega_{\lambda\nu}=0, & \\
	 \left(\varpi_{\lambda\nu}-\Omega_{\lambda\nu}\right)=0,
\end{cases}
\end{equation*}
which are both satisfied for $\Omega_{\mu\nu}=\varpi_{\mu\nu}$.
The form of the statistical operator at global equilibrium is then obtained by replacing these global equilibrium
forms for the thermal fields $\beta,\, \Omega$, and $\zeta$ in the statistical operator~\eqref{eq:rhoLTEphi}.
This form is the following:
\begin{equation*}
\begin{split}
\wrho\subs{GE}^{\,\Phi}=&\frac{1}{\mathcal{Z}}\exp\left\{-\int\!\!\D\Sigma_\mu\left[ \wT^{\mu\nu}\subs{B}\left(b_\nu + \varpi_{\nu\lambda}x^\lambda\right) -\wj^\mu\zeta\right]\right\}\\
	=&\frac{1}{\mathcal{Z}}\!\exp\!\left\{\!-b_\nu\!\!\!\int\!\!\!\D\Sigma_\mu \wT^{\mu\nu}\subs{B}\!\!-\! \varpi_{\nu\lambda}\!\!\int\!\!\!\D\Sigma_\mu \wT^{\mu\nu}\subs{B}\!x^\lambda
		\!+\!\zeta\!\!\int\!\!\!\D\Sigma_\mu\wj^\mu\!\!\right\}.
\end{split}
\end{equation*}
In the first and last terms of the last line, one can readily recognize the total momentum operator $\h{P}$ and the charge $\h{Q}$, respectively.
In the second term, by taking advantage of the anti-symmetric indices, the total boost-angular momentum $\h{J}$ is obtained.
The global equilibrium statistical operator is
\begin{equation}
\label{eq:rhoGE}
	\wrho\subs{GE}=\frac{1}{\mathcal{Z}}\exp\left[-b\cdot\wP+\frac{1}{2}\varpi:\wJ+\zeta\wQ\right].
\end{equation}
This derivation shows that no matter what the form of the pseudogauge fields $\wPhi$ and $\h{Z}$ is, the global equilibrium
statistical operator will always have the same form; therefore $\wrho\subs{GE}$ is pseudogauge independent.
Notice that if one chooses to constraint only the EMT but not the spin tensor, one would have to set $\Omega=0$. Therefore, when imposing the global
equilibrium, one would obtain only the particular case of the previous operator with $\varpi=0$. The only exceptions are the pseudogauges with
a vanishing spin current, such as the Belinfante one. In these cases, since all the spin information is contained in the EMT, one still obtains
the global equilibrium with non-vanishing thermal vorticity without introducing a spin potential~\cite{Buzzegoli:2020ycf}.

It follows that the PGTs do not affect the predictions of spin polarization and of the axial vortical effect (AVE) when the system is at global thermal
equilibrium. The spin polarization at the first order of thermal vorticity of a Dirac particle with momentum $k$ in a system at global thermal equilibrium
is
\begin{equation*}
S^\mu(k)= - \frac{1}{8m} \epsilon^{\mu\rho\sigma\tau} k_\tau \frac{\int_{\Sigma} \D \Sigma \cdot k \, n\subs{F} \left(1 -n\subs{F}\right) \varpi_{\rho\sigma}}
  {\int_{\Sigma} \D \Sigma \cdot k \, n\subs{F}},
\end{equation*}
which has the same form of Eq.~\eqref{eq:PolVortIntro}, the only difference being that at global equilibrium
the thermal vorticity must be a constant tensor. For the spin polarization at all orders of thermal vorticity see~\cite{Palermo:2021hlf}.

Similarly, for the Dirac field, the mean axial current resulting from the Zubarev statistical operator is discussed both for
massive~\cite{Buzzegoli:2017cqy,Prokhorov:2018bql,Buzzegoli:2021jeh} and massless fermions~\cite{Buzzegoli:2018wpy,Prokhorov:2018bql,Buzzegoli:2020ycf}.
The result is that the mean axial current is
\begin{equation}
\label{eq:AVEGE}
\mean{\wj^\mu\subs{A}}\subs{GE}=W^A w^\mu + \mathcal{O}\left(\varpi^2\right)
\end{equation}
where the four-vector $w$ is defined by
\begin{equation}
\label{eq:thermalrot}
w^\mu =-\frac{1}{2}\epsilon^{\mu\rho\sigma\lambda}\varpi_{\rho\sigma} u_\lambda=\frac{\omega^\mu}{T},
\end{equation}
which is the local rotation of the fluid,
$$
\omega^\mu=-\tfrac{1}{2}\epsilon^{\mu\nu\rho\sigma}\de_\nu u_\rho u_\sigma ,
$$
divided by the temperature.
The thermal coefficient $W^A$ is the AVE conductivity, and for a free Dirac field is given by~\cite{Buzzegoli:2018wpy,Flachi:2017vlp}
\begin{equation}
\label{eq:WA}
W\sups{A}=\!\int_0^\infty\!\! \frac{\D k}{2\pi^2\beta}\! \frac{\varepsilon_k^2+k^2}{\varepsilon_k}
	\left[n\subs{F}(\beta\varepsilon_k-\zeta)+n\subs{F}(\beta\varepsilon_k+\zeta)\right]
\end{equation}
with $\beta=\sqrt{\beta^2}$, $\varepsilon_k^2=\vec{k}^2+m^2$, and $m$ being the mass of the field.
For a massless field, the AVE conductivity above becomes
\begin{equation*}
W^A=\frac{T^3}{6}+\frac{\mu^2}{2\pi^2}T,
\end{equation*}
where $\mu=\zeta\, T$ is the chemical potential.
The mean axial current~\eqref{eq:AVEGE} and the conductivity~\eqref{eq:WA} were obtained with the global equilibrium statistical
operator~\eqref{eq:rhoGE} and do not depend on the pseudogauge.
In Sec.~\ref{sec:AxialC} I obtain how the AVE~\eqref{eq:AVEGE} is modified in a system at local thermal
equilibrium with a generic pseudogauge.

\section{Pseudogauge dependence of Wigner function}
\label{sec:Wigner}
The Wigner function extends the concept of the classical distribution function to a quantum system.
Like the distribution function, from the Wigner function one can obtain many relevant expectation values in statistical quantum field theory.
As it is generally expected that different distribution functions result in different averages, a difference in the Wigner function
may result in a variation in the final expectation values. In this section I derive the effect of the pseudogauge transformations (PGTs)
in the Wigner function of the free Dirac field. I will show that different choices of $\wPhi$ and $\h{Z}$ in~\eqref{eq:rhoLTEphi} affect
the Wigner function. In Secs.~\ref{sec:Polarization} and~\ref{sec:AxialC} I will show that indeed these different Wigner functions give different results
for the spin polarization and the mean axial current.

The covariant Wigner function of a free Dirac field is defined as~\cite{Groot}
\begin{equation}
\label{eq:WignerF}
W_{AB}(x,k)=\tr\left(\wrho\,\,\h{W}_{AB}(x,k)\right),
\end{equation}
where $A,\,B$ denote the spinorial indices, and $\h{W}$ denotes the Wigner operator,
\begin{equation*}
\h{W}_{AB}(x,k)=\!\!\int\!\frac{\D^4y}{(2\pi)^4}\E^{-\I k\cdot y}:\overline{\Psi}_B \left( x+\frac{y}{2}\right)\Psi_A\left( x-\frac{y}{2}\right): ,
\end{equation*}
and the colons ``$:\,:$'' denotes the normal ordering. The mean axial current can be obtained once the Wigner function is known 
by performing the integral
\begin{equation*}
\begin{split}
j^\mu\subs{A}(x)=&\mean{\h{j}^\mu\subs{A}(x)}=\int\D^4k\,\tr_4\left[\gamma^\mu\gamma^5 W(x,k)\right],
\end{split}
\end{equation*}
where $\tr_4$ indicates the trace over the spinorial indices.
Similarly, the mean spin polarization of a particle state with momentum $k$ is obtained through~\cite{Becattini:2020sww}
\begin{equation*}
S^\mu(k)=\frac{1}{2}\frac{\int_\Sigma \D\Sigma \cdot k\,\tr_4\left[\gamma^\mu\gamma^5
		W^+(x,k)\right]}{\int_\Sigma \D\Sigma \cdot k\, \tr_4\left[W^+(x,k)\right]},
\end{equation*}
where $W^+$ is the future time-like part (that is the particle part) of the Wigner function,
obtained by the decomposition
\begin{equation}
\label{eq:WignerDecomp}
\begin{split}
W(x,k)=& \theta(k^2)\theta(k^0)W^+(x,k) + \theta(k^2)\theta(-k^0)W^-(x,k)+\\
	& + \theta(-k^2)W^S(x,k).
\end{split}
\end{equation}

Notice that the Wigner function~\eqref{eq:WignerF} is affected by the statistical operator $\wrho$ that describes the thermal
state of the system. I am now going to plug the pseudogauge dependent statistical operator~\eqref{eq:rhoLTEphi} in 
Eq.~\eqref{eq:WignerF}. Then, I will quantify the pseudogauge dependence of the Wigner function using the linear response theory.

In this work I am going to consider the pseudogauge potentials $\wPhi$ given in Eqs.~(\ref{eq:PGPotentialsGamma}, \ref{eq:PGPotentialsDe}, \ref{eq:PGPotentialGLW},
and \ref{eq:PGPotentialHW}). All of these operators are bilinears in the Dirac fields. Therefore, recalling that
\begin{equation*}
\frac{\I}{2} \bar\Psi(x)\left({\stackrel{\rightarrow}{\partial}}\!\,^\mu-{\stackrel{\leftarrow}{\partial}}\!\,^\mu\right)\Psi(x)
	=\int\D^4 k\,k^\nu \h{W}(x,k),
\end{equation*}
they can be obtained from the Wigner operator through the integral
\begin{equation*}
	\wPhi^{\lambda,\mu\nu}(x)=\int\D^4k' \Phi^{\lambda,\mu\nu}_{B'A'}(k') \h{W}_{A'B'}(x,k'),
\end{equation*}
where $\Phi^{\lambda,\mu\nu}_{B'A'}$ is a function that might depend on the gamma matrices and on the metric tensors, but
that does not contain the derivative operator. See the Appendix \ref{sec:AppA} for the list of all the functions $\Phi$
used in this work. Among the gauge-potentials~(\ref{eq:PGPotentialsGamma} - \ref{eq:PGPotentialHW}) only the
Hilgevoord-Wouthuysen (HW) decomposition has a non-vanishing $\h{Z}$ potential. The pseudogauge-dependent statistical
operator~\eqref{eq:rhoLTEphi} only depends on the derivative of $\h{Z}$, which for the form in Eq.~\eqref{eq:PGPotentialHW}
can be written as
\begin{equation*}
\begin{split}
\de_\alpha\h{Z}^{\rho\sigma,\lambda\alpha}\subs{HW}(x)=& -\frac{1}{8m}\left(\sigma^{\rho\sigma}\sigma^{\lambda\alpha}+\sigma^{\lambda\alpha}\sigma^{\rho\sigma}\right)_{B'A'}\\
	&\times \int\D^4k'  \de_\alpha\h{W}_{A'B'}(x,k').
\end{split}
\end{equation*}
Lastly, the improved Callan, Coleman, and Jackiew (CCJ) pseudogauge potentials~\eqref{eq:PGPotentialCCJ}
can be obtained from the Wigner operator by
\begin{equation}
\label{eq:CCJPotentials}
\begin{split}
\wPhi^{\lambda,\mu\nu}\subs{CCJ}(x)=&-\frac{1}{6}\left(\eta^{\lambda\nu}\partial^\mu-\eta^{\lambda\mu}\partial^\nu\right)\int\D^4k' \tr_4 \left(\h{W}(x,k')\right),\\
\de_\alpha\h{Z}^{\rho\sigma,\lambda\alpha}\subs{CCJ}=&-\frac{1}{6}\left(\eta^{\alpha\rho}\eta^{\lambda\sigma}-\eta^{\alpha\sigma}\eta^{\lambda\rho}\right)\\
	&\times \int\D^4k'  \de_\alpha\tr_4 \left(\h{W}(x,k')\right).
\end{split}
\end{equation}
However, eventually it is found that the pseudogauge potentials $\wPhi\subs{CCJ}$ and $\h{Z}\subs{CCJ}$ in Eq.~\eqref{eq:PGPotentialCCJ} do not affect the
Wigner function at first order in linear response theory.

As mentioned before, the pseudogauge-dependent Wigner function at local thermal equilibrium is obtained by
\begin{equation*}
\mean{\h{W}_{AB}(x,k)}_\Phi=\tr\left(\wrho\subs{LTE}^{\,\Phi}\,\,\h{W}_{AB}(x,k)\right),
\end{equation*}
where, setting $\zeta=0$ for clarity, and introducing the shorthand notation
\begin{equation}
\label{eq:ThetaDef}
	\Theta_{\mu\nu}\equiv \varpi_{\mu\nu}-\Omega_{\mu\nu},
\end{equation}
one has
\begin{equation}
\label{eq:rhoPhi}
\begin{split}
\wrho\subs{LTE}^{\,\Phi}=&\frac{1}{\mathcal{Z}}\!\exp\!\left\{\!\!-\!\!\int\!\!\D\Sigma_\lambda\!\!\left[ \wT^{\lambda\nu}\subs{B}\beta_\nu
	-\frac{1}{2}\Theta_{\rho\sigma}\wPhi^{\lambda,\rho\sigma} - \xi_{\rho\sigma} \wPhi^{\rho,\lambda\sigma}+\right.\right.\\
	&\left.\left.-\frac{1}{2}\Omega_{\rho\sigma}\de_\alpha\h{Z}^{\rho\sigma,\lambda\alpha}\right]\right\},
\end{split}
\end{equation}
and the operators $\wPhi$ and $\de_\alpha\h{Z}^{\rho\sigma,\lambda\alpha}$ can be written as described above. For a fluid
in the hydrodynamic regime, such as the QGP, the thermodynamics field $\beta$ slowly varies compared to the correlation lengths
between two operators. Furthermore, since no effect of the spin potential $\Omega$ was ever observed, one assumes that it
is not larger than the thermal vorticity $\varpi$. Therefore the terms in the exponent of~\eqref{eq:rhoPhi} which are
coupled to $\Theta,\,\xi$, and $\Omega$ can be treated as ``small'' perturbations.

One should now study the difference between the Wigner function resulting from a generic pseudogauge and the one obtained in
the Belinfante decomposition, which is denoted as
\begin{equation*}
\mean{\h{W}_{AB}(x,k)}\subs{B}=\tr\left(\wrho\subs{B}^{\,\Phi}\,\,\h{W}_{AB}(x,k)\right).
\end{equation*}
For the sake of clarity, I report only the steps for the particle part of the Wigner function $W^+$; see Eq.~\eqref{eq:WignerDecomp}.
The other parts are carried out in a similar fashion. Denoting the thermal average of an operator $\h{O}$ made with the statistical operator
\begin{equation*}
\wrho_\beta= \frac{1}{\mathcal{Z}}\exp\left\{-\beta_\nu(x)\int\!\!\D\Sigma_\lambda \, \wT^{\lambda\nu}\subs{B}\right\}
	= \frac{1}{\mathcal{Z}}\exp\left\{-\beta(x)\cdot \h{P}\right\}
\end{equation*}
as $\mean{\h{O}}_{\beta(x)}=\tr (\wrho_\beta\,\h{O} )$ and using the linear response theory as described
in~\cite{vanWeert,Becattini:2015nva,Buzzegoli:2020ycf,Becattini:2021suc}, one obtains:
\begin{equation*}
\begin{split}
\Delta_\Phi& W^+_{AB}(x,k)= \mean{\h{W}^+_{AB}(x,k)}_\Phi - \mean{\h{W}^+_{AB}(x,k)}\subs{B}\\
	&\simeq\Delta_\Theta W^+_{AB}(x,k) + \Delta_\xi W^+_{AB}(x,k)+ \Delta_{\Omega Z} W^+_{AB}(x,k),
\end{split}
\end{equation*}
where
\begin{widetext}
\begin{equation*}
\begin{split}
\Delta_\Theta W^+_{AB}(x,k)=& \frac{1}{2}\int_0^1\D z\int_{\Sigma}\D\Sigma_\lambda(y)\Theta_{\rho\sigma}(y) \int\D^4 k' \Phi^{\lambda,\rho\sigma}_{B'A'}
	\mean{\h{W}^+_{AB}(x,k)\h{W}_{A'B'}(y+\I z\beta(x),k')}_{\beta(x)},\\
\Delta_\xi W^+_{AB}(x,k)=&\int_0^1\D z\int_{\Sigma}\D\Sigma_\lambda(y)\xi_{\rho\sigma}(y)\int\D^4 k'
	\Phi^{\rho,\lambda\sigma}_{B'A'}\mean{\h{W}^+_{AB}(x,k)\h{W}_{A'B'}(y+\I z\beta(x),k')}_{\beta(x)},\\
\Delta_{\Omega Z} W^+_{AB}(x,k)=& \frac{1}{2}\int_0^1\D z\int_{\Sigma}\D\Sigma_\lambda(y)\Theta_{\rho\sigma}(y) \int\D^4 k' Z^{\rho\sigma,\lambda\alpha}_{B'A'}
	{\de_y}_\alpha\mean{\h{W}^+_{AB}(x,k)\h{W}_{A'B'}(y+\I z\beta(x),k')}_{\beta(x)}.
\end{split}
\end{equation*}

Using standard thermal field theory techniques (see for instance~\cite{Becattini:2020xbh,Becattini:2021suc}), one finds
\begin{equation*}
\begin{split}
\mean{\h{W}^+_{ab}(x,k)\h{W}_{cd}(y+\I z\beta(x),k')}_{c,\beta(x)}= &
	\frac{1}{(2\pi)^6}\int\frac{\D^3\p}{2\varepsilon_p}\int\frac{\D^3\p'}{2\varepsilon_{p'}}
	\delta^4\left(k-\frac{p+p'}{2} \right)\delta^4\left(k'-\frac{p+p'}{2} \right)\\
& \times(\slashed{p}'+m)_{ad}(\slashed{p}+m)_{cb}
	\E^{\I(p-p')(x-y)}\E^{z(p-p')\beta}n\subs{F}(\beta(x)\cdot p)\left(1-n\subs{F}(\beta(x)\cdot p')\right)
\end{split}
\end{equation*}
from which it straightforwardly follows that
\begin{equation}
\label{eq:DeltaW}
\begin{split}
\Delta_\Theta W^+_{AB}(x,k)=& \frac{1}{2}\int_0^1\D z\int_{\Sigma}\D\Sigma_\lambda(y)\Theta_{\rho\sigma}(y)
	\frac{1}{(2\pi)^6}\int\frac{\D^3\p}{2\varepsilon_p}\int\frac{\D^3\p'}{2\varepsilon_{p'}}\delta^4\left(k-\frac{p+p'}{2} \right)\\
	&\times \left[(\slashed{p}'+m)\Phi^{\lambda,\rho\sigma}(\slashed{p}+m)\right]_{AB}
	\E^{\I(p-p')(x-y)}\E^{z(p-p')\beta}n\subs{F}(\beta(x)\cdot p)\left(1-n\subs{F}(\beta(x)\cdot p')\right),\\
\Delta_\xi W^+_{AB}(x,k)=&\int_0^1\D z\int_{\Sigma}\D\Sigma_\lambda(y)\xi_{\rho\sigma}(y)
	\frac{1}{(2\pi)^6}\int\frac{\D^3\p}{2\varepsilon_p}\int\frac{\D^3\p'}{2\varepsilon_{p'}}\delta^4\left(k-\frac{p+p'}{2} \right)\\
	&\times \left[(\slashed{p}'+m)\Phi^{\rho,\lambda\sigma}(\slashed{p}+m)\right]_{AB}
	\E^{\I(p-p')(x-y)}\E^{z(p-p')\beta}n\subs{F}(\beta(x)\cdot p)\left(1-n\subs{F}(\beta(x)\cdot p')\right),\\
\Delta_{\Omega Z} W^+_{AB}(x,k)=& -\frac{\I}{2}\int_0^1\D z\int_{\Sigma}\D\Sigma_\lambda(y)\Omega_{\rho\sigma}(y)
	\frac{1}{(2\pi)^6}\int\frac{\D^3\p}{2\varepsilon_p}\int\frac{\D^3\p'}{2\varepsilon_{p'}}\delta^4\left(k-\frac{p+p'}{2} \right)\\
	&\times \left[(\slashed{p}'+m)Z^{\rho\sigma,\lambda\alpha}(\slashed{p}+m)\right]_{AB}(p-p')_\alpha
	\E^{\I(p-p')(x-y)}\E^{z(p-p')\beta}n\subs{F}(\beta(x)\cdot p)\left(1-n\subs{F}(\beta(x)\cdot p')\right).
\end{split}
\end{equation}
\end{widetext}

Notice that the integration over the space-like hypersurface $\Sigma$, i.e., in the $y$ coordinates, only involves
the exponential $\E^{\I(p-p')(x-y)}$ and the thermodynamic fields $\Theta$, $\xi$, and $\Omega$. As mentioned
above, in the hydrodynamic regime the thermodynamic fields are slowly varying, and the integral over $\Sigma$ 
can be approximated by Taylor expanding the thermodynamic fields around the point $y=x$ and retaining only
the leading order. For instance, the expansion of $\Theta$ is
\begin{equation*}
	\Theta_{\rho\sigma}(y)=\Theta_{\rho\sigma}(x)+\de_\kappa \Theta_{\rho\sigma}(x) (y-x)^\kappa+\cdots,
\end{equation*}
and one obtains (see~\cite{Becattini:2020xbh,Becattini:2021suc})
\begin{equation}
\label{eq:IntSigma}
\begin{split}
	\int_{\Sigma}\!\!\!\D\Sigma_\lambda(y)\Theta_{\rho\sigma}(y) \E^{\I(p-p')(x-y)} \simeq
	\Theta_{\rho\sigma}(x)\hat{t}_\lambda (2\pi)^3\delta^3(\vec{p}\!-\!\vec{p'}),
\end{split}
\end{equation}
where $\hat{t}$ is the time direction in the fluid frame. After this approximation, thanks to the delta function,
it is straightforward to integrate the~\eqref{eq:DeltaW} in $p'$. As a result, $\Delta_{\Omega Z} W^+$ is vanishing
because of the factor $p-p'$, while in the other contributions the dependence on $z$ goes away. Lastly, taking advantage
of the transformation
\begin{equation*}
\begin{split}
\int\!\!\frac{\D^3 p}{2\epsilon_p}\delta^4(k-p)f(p)=&
	\!\!\int\!\!\D^4 k \delta(k^2\!\!-m^2)\theta(k_0)\delta^4(k\!-\! p)f(p)\\
=&\theta(k_0)\delta(k^2-m^2)f(k),
\end{split}
\end{equation*}
these final expressions are obtained:
\begin{widetext}
\begin{equation}
\label{eq:DeltaWigPseudoFinal}
\begin{split}
\Delta_\Theta W^+_{AB}(x,k)=& \frac{\theta(k_0)\delta(k^2-m^2)}{4(2\pi)^3} \frac{\hat{t}_\lambda\Theta_{\rho\sigma}(x)}{\epsilon_k}
	\left[(\slashed{k}+m)\Phi^{\lambda,\rho\sigma}(\slashed{k}+m)\right]_{AB} n\subs{F}(\beta(x)\cdot k)\left(1-n\subs{F}(\beta(x)\cdot k)\right),\\
\Delta_\xi W^+_{AB}(x,k)=&\frac{\theta(k_0)\delta(k^2-m^2)}{2(2\pi)^3} \frac{\hat{t}_\lambda\xi_{\rho\sigma}(x)}{\epsilon_k}
	\left[(\slashed{k}+m)\Phi^{\rho,\lambda\sigma}(\slashed{k}+m)\right]_{AB} n\subs{F}(\beta(x)\cdot k)\left(1-n\subs{F}(\beta(x)\cdot k)\right),\\
\Delta_{\Omega Z} W^+_{AB}(x,k)=& 0.
\end{split}
\end{equation}
\end{widetext}
Equation~\eqref{eq:DeltaWigPseudoFinal} clearly shows that the Wigner function of local thermal equilibrium does depend on the choice of pseudogauge potentials.
In some cases, these corrections might be vanishing. For instance, if $\wPhi^{\rho,\lambda\sigma}$ is completely anti-symmetric,
such as in $\wPhi\subs{C},\,\wPhi_{\epsilon{\rm A}},\,\wPhi_{\epsilon{\rm V}}$, $\wPhi_{\epsilon\de},\,\wPhi_{\epsilon\de{\rm A}}$,
then one readily has that $\Delta_\xi W^+_{AB}(x,k)=0$. One could also repeat the argument above for the improved Callan, Coleman, and Jackiew
pseudogauge potentials~\eqref{eq:PGPotentialCCJ} using~\eqref{eq:CCJPotentials}. As in $\Delta_{\Omega Z} W^+$, the derivatives
of Wigner function in~\eqref{eq:CCJPotentials} create factors $p-p'$ which vanish when integrating the Dirac delta from the
approximation~\eqref{eq:IntSigma}. Therefore, at first order in linear perturbation theory, the Callan, Coleman, and Jackiew decomposition
gives the same thermal expectation values as the Belinfante decomposition. In the next sections I show that this difference in the Wigner
function actually affects physical observables.

\section{Pseudogauge dependence of spin polarization}
\label{sec:Polarization}
The spin polarization is obtained from the Wigner function by~\cite{Becattini:2020sww}
\begin{equation*}
S^\mu(k)=\frac{1}{2}\frac{\int_\Sigma \D\Sigma \cdot k\,\tr_4\left[\gamma^\mu\gamma^5
		W^+(x,k)\right]}{\int_\Sigma \D\Sigma_\alpha  k^\alpha \tr_4\left[W^+(x,k)\right]}.
\end{equation*}
It is important to stress that the form of this formula does not depend on the pseudogauge. The pseudogauge
dependence of spin polarization from the previous formula is completely contained inside the Wigner function, which
has itself inherited the pseudogauge dependence from the statistical operator. As mentioned, the spin polarization
in the Belinfante PG~\eqref{eq:rhoLTEBintro} is given by
\begin{equation}
\label{eq:PolBel}
S^\mu\subs{B}(k)\simeq S^\mu_{\varpi}(k) + S^\mu_{\xi}(k);
\end{equation}
see Eqs.~\eqref{eq:PolVortIntro} and~\eqref{eq:PolShearIntro}. Under the hypothesis that linear response theory is a
good approximation, the difference between the spin polarization in a generic pseudogauge $S^\mu_\Phi(k)$ and
in the Belinfante $S^\mu\subs{B}(k)$ can be obtained from the results of the previous section.
Using the above formula, it is given by:
\begin{equation}
\label{eq:DeltaPol}
\begin{split}
\Delta_\Phi S^\mu(k)=& S^\mu_\Phi(k) - S^\mu\subs{B}(k)\\
 = & \frac{1}{2{\cal D}}\int_\Sigma \D\Sigma \cdot k\,\tr_4\left[\gamma^\mu\gamma^5
		\Delta_\Phi W^+(x,k)\right],
\end{split}
\end{equation}
with
\begin{equation*}
\begin{split}
{\cal D}=& \int_\Sigma \D\Sigma_\alpha  k^\alpha \tr_4\left[\mean{\h{W}^+(x,k)}_{\beta(x)}\right]\\
=&\frac{4m}{(2\pi)^3}\int_\Sigma\!\! \D\Sigma \cdot k \; \delta(k^2-m^2)\theta(k_0) n\subs{F},
\end{split}
\end{equation*}
where $n\subs{F}=n\subs{F}(\beta(x)\cdot k)$. Plugging in the results of~\eqref{eq:DeltaWigPseudoFinal} into~\eqref{eq:DeltaPol},
we obtain
\begin{equation}
\label{eq:DeltaS}
\begin{split}
\Delta_\Phi S^\mu(k)=&\Delta_\Theta S^\mu(k)+\Delta_\xi S^\mu(k),\\
\Delta_\Theta S^\mu(k)=&\frac{A^{\mu\rho\sigma\lambda}_{\Theta,\,\Phi}\hat{t}_\lambda}{32 m\varepsilon_k}
	 \frac{\int_\Sigma \D\Sigma\cdot k\, \Theta_{\rho\sigma} n\subs{F}\left(1-n\subs{F}\right)}{\int_\Sigma \D\Sigma \cdot k \, n\subs{F}}, \\
\Delta_\xi S^\mu(k)=&\frac{A^{\mu\rho\sigma\lambda}_{\xi,\,\Phi}\hat{t}_\lambda}{16 m\varepsilon_k}
	 \frac{\int_\Sigma \D\Sigma\cdot k\, \xi_{\rho\sigma} n\subs{F}\left(1-n\subs{F}\right)}{\int_\Sigma \D\Sigma \cdot k \, n\subs{F}},
\end{split}
\end{equation}
where
\begin{equation*}
\begin{split}
A^{\mu\rho\sigma\lambda}_{\Theta,\,\Phi}=&\tr_4\left[\gamma^\mu\gamma^5(\slashed{k}+m)\Phi^{\lambda,\rho\sigma}(\slashed{k}+m) \right], \\
A^{\mu\rho\sigma\lambda}_{\xi,\,\Phi}=&\tr_4\left[\gamma^\mu\gamma^5(\slashed{k}+m)\Phi^{\rho,\lambda\sigma}(\slashed{k}+m) \right].
\end{split}
\end{equation*}
These are ordinary gamma matrix traces and result in
\begin{widetext}
\begin{equation}\label{eq:ATheta}
A^{\mu\rho\sigma\lambda}_{\Theta,\,\Phi}=\begin{cases}
4\epsilon^{\lambda\rho\sigma\tau}\left(k^\mu k_\tau-\eta^\mu_{\,\tau}m^2\right), & \Phi\subs{C}=\Phi_{\epsilon {\rm A}}, \\
16\left(k^\mu k^\sigma \eta^{\lambda\rho}+m^2\eta^{\mu\rho}\eta^{\lambda\sigma}\right), & \Phi\subs{A}, \\
-16 \epsilon^{\mu\rho\sigma\tau} k_\tau k^\lambda, & \Phi_{\de\Sigma}, \\
A^{\mu\rho\sigma\lambda}_{\Theta,\,\Phi\subs{C}}+ 8\epsilon^{\lambda\mu\rho\tau} k_\tau k^\sigma, &  \Phi\subs{GLW}\text{ and } \Phi\subs{HW},
\end{cases}
\end{equation}
and
\begin{equation}\label{eq:AXi}
A^{\mu\rho\sigma\lambda}_{\xi,\,\Phi}=\begin{cases}
0, & \Phi\subs{C}=\Phi_{\epsilon {\rm A}}, \\
8\left(k^\mu k^\sigma-m^2\eta^{\mu\sigma}\right) \eta^{\lambda\rho} -8\left(k^\mu k^\lambda - m^2\eta^{\mu\lambda}\right)\eta^{\rho\sigma}, & \Phi\subs{A}, \\
16 \epsilon^{\lambda\mu\sigma\tau} k_\tau k^\rho, & \Phi_{\de\Sigma}, \\
-4\epsilon^{\lambda\mu\sigma\tau} k_\tau k^\rho, &  \Phi\subs{GLW}\text{ and } \Phi\subs{HW},
\end{cases}
\end{equation}
and are vanishing for the other pseudogauges considered in this work,~(\ref{eq:PGPotentialsGamma} - \ref{eq:PGPotentialCCJ}).
\end{widetext}
\subsection{Canonical decomposition}
\label{subsec:CanonicalPol}
It is worth discussing some special cases of the general result~\eqref{eq:DeltaS}.
If one chooses to describe the local thermodynamic equilibrium with the canonical decomposition, one should use
the statistical operator
\begin{equation*}
\wrho\subs{LTE}\sups{C}=\frac{1}{\mathcal{Z}}\exp\left[-\int\D\Sigma_\mu\!\left( \wT^{\mu\nu}\subs{C}\beta_\nu
	\!-\frac{1}{2}\Omega_{\lambda\nu}\wspt\subs{C}^{\mu,\lambda\nu}\right)\right],
\end{equation*}
which, as shown in Sec.~\ref{sec:LocalEquilibrium}, can be rewritten by means of the pseudogauge
transformation~\eqref{eq:PsGaugeBel} with
\begin{equation*}
\wPhi^{\lambda,\mu\nu}\subs{C}=- \wspt\subs{C}^{\lambda,\mu\nu}
		=-\frac{\I}{8}\bar\Psi\left\{\gamma^\lambda,\left[\gamma^\mu,\gamma^\nu\right]\right\}\Psi
\end{equation*}
and $\h{Z}\subs{C}=0$, as\footnote{For the Dirac field, the canonical spin tensor is completely anti-symmetric and the shear term in \eqref{eq:rhoLTEphi} vanishes.}
\begin{equation}
\label{eq:LTECan}
\wrho\subs{LTE}\sups{C}=\frac{1}{\mathcal{Z}}\exp\left[-\int\D\Sigma_\mu\!\left( \wT^{\mu\nu}\subs{B}\beta_\nu
	\!+\frac{\varpi_{\lambda\nu}-\Omega_{\lambda\nu}}{2}\wspt\subs{C}^{\lambda,\mu\nu}\right)\right].
\end{equation}
The spin polarization resulting from this operator is
\begin{equation}
\label{eq:PolCan}
S^\mu\subs{C}(k)\simeq S^\mu_{\varpi}(k) + S^\mu_{\xi}(k) + \Delta\sups{C}_\Theta S^\mu(k).
\end{equation}
The first two terms are given by Eq.s~\eqref{eq:PolVortIntro} and~\eqref{eq:PolShearIntro} and
are the contributions coming from the Belinfante form, which appears in the first term in the exponent
of Eq.~\eqref{eq:LTECan}. These two are the terms that have been considered so far to predict the spin polarization of particles emitted by the quark gluon plasma.
When adopting the canonical decomposition, in addition to these terms we should also add the contribution from the canonical spin tensor $\Delta\sups{C}_\Theta S^\mu(k)$.
This one can be read from~\eqref{eq:DeltaS} and it is given by
\begin{equation}
\label{eq:DeltaSCan}
\begin{split}
\Delta\sups{C}_\Theta S^\mu(k)=&\frac{\epsilon^{\lambda\rho\sigma\tau}\hat{t}_\lambda(k^\mu k_\tau-\eta^\mu_{\;\tau}m^2)}{8 m \varepsilon_k}\\
	&\times \frac{\int_\Sigma \D\Sigma(x)\cdot k\, n\subs{F}\left(1-n\subs{F}\right)\left(\varpi_{\rho\sigma}-\Omega_{\rho\sigma}\right)}{\int_\Sigma \D\Sigma \cdot k \, n\subs{F}}.
\end{split}
\end{equation}
To the best of my knowledge, this is the first time that an explicit formula for the contribution of (canonical) spin tensor is reported.
Notice that in this case there is no additional contribution from the thermal shear $\xi$, as the canonical spin tensor is completely
anti-symmetric.

The contribution to spin polarization in Eq.~\eqref{eq:DeltaSCan} is the first important and necessary step to study the impact of pseudogauge
transformations to the spin polarization in heavy-ion collisions. For instance, the result~\eqref{eq:DeltaSCan} reveals that the contribution
of spin potential is not simply obtained by replacing the thermal vorticity with the spin potential in Eq.~\eqref{eq:PolVortIntro}.
An other important conclusion that one draws from Eq.~\eqref{eq:DeltaSCan} is that it is impossible to remove the difference
between the spin polarization in the Belinfante and canonical PGs by choosing different values of the $\beta(x)$ and
$\Omega(x)$ fields for the two PGs. Indeed, since the $k$ dependence  makes the Eqs.~\eqref{eq:PolVortIntro}, \eqref{eq:PolShearIntro},
and~\eqref{eq:DeltaSCan} linearly independent, the only way to satisfy the equivalence
$S^\mu\subs{B}[\beta^{\rm B}](k)=S^\mu\subs{C}[\beta^{\rm C},\Omega^{\rm C}](k)$ for all the values of $k$, is to choose $\beta^{\rm B}=\beta^{\rm C}$
and $\Omega^{\rm C}=\varpi^{\rm C}$, that is the same $\beta$ field for both the PGs and the condition for global equilibrium.

Clearly, a quantitative estimate requires a numerical analysis.  To assess the relevance of the contribution~\eqref{eq:DeltaSCan} one should know the
magnitude of the difference between the thermal vorticity and the spin potential. While one can evaluate the thermal vorticity, for instance with
hydrodynamic simulations, the magnitude of the spin potential $\Omega$, which appears as a parameter to be fixed with observation, is not known.
However, as discussed in Sec.~\ref{subsec:global}, the quantity $\Theta=\varpi-\Omega$ must reach zero as the system approaches global
equilibrium and it is therefore expected to be significant only for systems far from equilibrium. Also note that the manifest breaking of Lorentz
covariance in~\eqref{eq:DeltaSCan} by the presence of the unit vector $\hat{t}$ is expected as the canonical spin tensor is not a conserved quantity
and the statistical operator~\eqref{eq:LTECan} depends on the particular 3D hypersurface of integration; see also the discussion
in~\cite{Becattini:2021suc}.

\subsection{de Groot-van Leeuwen-van Weert and Hilgevoord-Wouthuysen decompositions}
Consider now the de Groot-van Leeuwen-van Weert (GLW) and Hilgevoord-Wouthuysen (HW) decompositions.
Both these decompositions result in the same predictions for the spin polarization. From the
results~\eqref{eq:DeltaS} the spin polarization reads:
\begin{equation*}
\begin{split}
S^\mu\subs{GLW,HW}(k)\simeq& S^\mu_{\varpi}(k) + S^\mu_{\xi}(k) + \Delta\sups{C}_\Theta S^\mu(k)+\\
 &+\Delta\sups{GLW,HW}_\Theta S^\mu(k) 	+ \Delta\sups{GLW,HW}_\xi S^\mu(k).
\end{split}
\end{equation*}
The first three terms were discussed above, the other two are given by:
\begin{equation}
\label{eq:DeltaThetaGLW}
\begin{split}
\Delta\sups{GLW,HW}_\Theta S^\mu(k) = & -\frac{1}{4m}\epsilon^{\mu\lambda\rho\tau}\hat{t}_\lambda\frac{ k_\tau k^\sigma }{\varepsilon_k}\\
	&\times \frac{\int_\Sigma \D\Sigma\cdot k\, n\subs{F}\left(1-n\subs{F}\right) \left(\varpi_{\rho\sigma}-\Omega_{\rho\sigma}\right)}{\int_\Sigma \D\Sigma \cdot k \, n\subs{F}},
\end{split}
\end{equation}
and
\begin{equation}
\label{eq:DeltaXiGLW}
\begin{split}
\Delta\sups{GLW,HW}_\xi S^\mu(k) = &-S_\xi^\mu(k)=+\frac{1}{4m}\epsilon^{\mu\lambda\sigma\tau}\frac{ k_\tau k^\rho }{\varepsilon_k}\\
	&\times \frac{\int_\Sigma \D\Sigma\cdot k\, n\subs{F}\left(1-n\subs{F}\right)\hat{t}_\lambda\xi_{\rho\sigma}}{\int_\Sigma \D\Sigma \cdot k \, n\subs{F}}.
\end{split}
\end{equation}
The contribution in Eq.~\eqref{eq:DeltaThetaGLW} has a different form compared to both Eq.~\eqref{eq:PolVortIntro}
and Eq.~\eqref{eq:DeltaSCan}, and its numerical investigation requires the knowledge of the spin tensor $\Omega$.
Instead, the contribution of~\eqref{eq:DeltaXiGLW} only requires the knowledge of the thermal shear $\xi$ and it must
be included even if the spin potential is not introduced at all. Remarkably, it cancels exactly the contribution of
Eq.~\eqref{eq:PolShearIntro}. The remaining terms can be simplified using the Schouten identity, obtaining the more compact formula:
\begin{equation*}
\begin{split}
S^\mu\subs{GLW,HW}(k)=&- \frac{1}{8m} \epsilon^{\mu\rho\sigma\tau} k_\tau \frac{\int_{\Sigma} \D \Sigma \cdot k \, n\subs{F} \left(1 -n\subs{F}\right) \Omega_{\rho\sigma}}
  {\int_{\Sigma} \D \Sigma \cdot k \, n\subs{F}}.
\end{split}
\end{equation*}
As found for the canonical PG, even allowing for different values of the thermodynamic fields, the only way
to obtain the same predictions for the spin polarization in the HW/GLW PG as in the Belinfante or canonical PG is to have
the same $\beta$ field in all PGs and to impose the global equilibrium conditions $\xi=0$ and $\Omega=\varpi$.

The importance of including the thermal shear term in Eq.~\eqref{eq:PolShearIntro} in heavy-ion collisions has been
discussed in~\cite{Becattini:2021iol,Fu:2021pok}, where it was found that such a term is able to restore the
agreement between the predictions and the experimental data for the momentum dependent spin polarization of
Lambda hyperons. The above studies were (tacitly) carried out in the Belinfante decomposition.
From the analysis above, if the GLW or HW decomposition were used instead,
there would have been no thermal-shear corrections and hence no agreement with the data (with the predictions of local
thermal equilibrium at first order in the gradients).
This shows how the choice of the pseudogauge can indeed have sizable effects on observables.

\subsection{Other pseudogauges}
\label{subsec:OtherPGs}
In \emph{special} relativity, the choice of the pseudogauge is arbitrary. In this work, I considered only a few choices
for the pseudogauges of the Dirac field, see Eq.s~(\ref{eq:PGPotentialsGamma} - \ref{eq:PGPotentialCCJ}), and I evaluated
the resulting predictions for the spin polarization~\eqref{eq:DeltaS}. It was found that in the GLW and HW decompositions,
the contribution of the thermal shear is increased compared to the Belinfante. If instead one chooses the $\Phi_{\de\Sigma}$
pseudogauge in Eq.~\eqref{eq:PGPotentialsDe}, one would obtain
\begin{equation}
\label{eq:PolDeSigma}
\Delta^{\de\Sigma}_\xi S^\mu(k)=\frac{-1}{m}\epsilon^{\mu\lambda\sigma\tau}\frac{ k_\tau k^\rho }{\varepsilon_k}
	 \frac{\int_\Sigma \D\Sigma\cdot k\, n\subs{F}\left(1-n\subs{F}\right)\hat{t}_\lambda\xi_{\rho\sigma}}{\int_\Sigma \D\Sigma \cdot k \, n\subs{F}},
\end{equation}
which is larger compared to the contribution~\eqref{eq:PolShearIntro}. Adding the two terms \eqref{eq:PolDeSigma}
and \eqref{eq:PolShearIntro} together will enhance the effects discussed in~\cite{Becattini:2021iol}. However, both
the sign and the numerical factor of the pseudogauge $\Phi_{\de\Sigma}$ are arbitrary. Indeed, in special relativity, one could
have used any linear combinations of the pseudogauges in~(\ref{eq:PGPotentialsGamma} - \ref{eq:PGPotentialCCJ}). It follows that
it should be easy to pick an \textit{ad-hoc} pseudogauge which is able to explain a certain set of data.
Therefore, such an approach should be avoided unless one is dealing with a large set
of measurements. In the lack of experimental evidence, one can also look at theoretical arguments that favor some of the pseudogauges.

The canonical, Belinfante, GLW and HW pseudogauges do not constitute an arbitrary or \textit{a posteriori} choice, as they
are the result of the direct application of Noether theorem or a specific choice in the symmetries and properties of the EMT
tensor or the spin tensor. When we have a strong case for using a specific pseudogauge, then we can use measurements made at local
equilibrium, such as the spin polarization in heavy-ion collisions, to discern what pseudogauge better describes the system.

\section{Pseudogauge dependence of mean axial current}
\label{sec:AxialC}
Let us now turn to the mean axial current, which is obtained from the Wigner function with
\begin{equation*}
j^\mu\subs{A}(x)=\mean{\h{j}^\mu\subs{A}(x)}=\int\D^4k\,\tr_4\left[\gamma^\mu\gamma^5 W(x,k)\right].
\end{equation*}
As previously discussed, at global equilibrium one obtains the axial current in Eq.~\eqref{eq:AVEGE}.
Instead, the mean axial current evaluated using the local equilibrium statistical operator in the Belinfante pseudogauge~\eqref{eq:rhoLTEB}
is
\begin{equation}
\label{eq:AXBel}
\mean{\wj^\mu\subs{A}}\subs{B}=W^A w^\mu + \epsilon^{\lambda\mu\sigma\tau}\xi_{\rho\sigma}\h{t}_\lambda \Xi_\tau^{\,\rho}
	+\mathcal{O}\left(\de\beta^2\right),
\end{equation}
where $W^A$ and $w$ are defined as in Eqs.~\eqref{eq:WA} and~\eqref{eq:thermalrot}, except that the thermal vorticity does not have
to be constant anymore. The second term in~\eqref{eq:AXBel} is the axial current induced by the thermal shear $\xi$, and
from the Wigner function evaluated in~\cite{Becattini:2021suc} one can show that
\begin{equation}
\label{eq:AXShearBel}
\Xi_\tau^{\,\rho} = 4\int\frac{\D^4 k}{(2\pi)^3\varepsilon_k}\theta(k^0)\delta(k^2-m^2)
	n\subs{F}(1-n\subs{F}) k_\tau k^\rho \, .
\end{equation}
However, the above formula must be modified in other pseudogauges. In a different pseudogauge one should add the difference:
\begin{equation*}
\Delta_\phi j^\mu\subs{A}(x) =  \mean{\h{j}^\mu\subs{A}(x)}_\Phi - \mean{\h{j}^\mu\subs{A}(x)}\subs{B}
	\simeq \Delta_\Theta j^\mu\subs{A}(x) + \Delta_\xi j^\mu\subs{A}(x).
\end{equation*}
Using the results in Eq.~\eqref{eq:DeltaWigPseudoFinal} for the particle part and similar expressions for the
anti-particle, one obtains
\begin{equation}
\label{eq:AxialPGDep}
\begin{split}
\Delta_\Theta j^\mu\subs{A}(x)=&\Theta_{\rho\sigma}\hat{t}_\lambda\int\frac{\D^4 k}{2(2\pi)^3\varepsilon_k}\theta(k^0)\delta(k^2-m^2)\\
	&\times n\subs{F}(1-n\subs{F})A^{\mu\rho\sigma\lambda}_{\Theta,\,\Phi}, \\
\Delta_\xi j^\mu\subs{A}(x)=&+\xi_{\rho\sigma}\hat{t}_\lambda\int\frac{\D^4 k}{(2\pi)^3\varepsilon_k}\theta(k^0)\delta(k^2-m^2)\\
	&\times n\subs{F}(1-n\subs{F})A^{\mu\rho\sigma\lambda}_{\xi,\,\Phi},
\end{split}
\end{equation}
where the quantities $A$ are the same as Eqs.~\eqref{eq:ATheta} and~\eqref{eq:AXi}. These modifications are generally
non-vanishing. In particular, the thermal field $\Theta$ is proportional to the thermal vorticity, so $\Delta_\Theta j^\mu\subs{A}$
contains a contribution along the rotation of the fluid, affecting the axial vortical effect conductivity.

As a notable example consider a massless field and the canonical decomposition where $\wPhi=\wPhi\subs{C}$, that is
\begin{equation*}
\begin{split}
\Delta\sups{C}_\Theta j^\mu\subs{A}(x)=& \epsilon^{\lambda\rho\sigma\tau} \hat{t}_\lambda \Theta_{\rho\sigma}
		\int\frac{\D^4 k}{4\pi^3 \varepsilon_k}\theta(k_0)\delta(k^2) k^\mu k_\tau\\
	&\times n\subs{F}(\beta(x)\cdot k)\left(1-n\subs{F}(\beta(x)\cdot k)\right)\\
=& \epsilon^{\lambda\rho\sigma\tau} \hat{t}_\lambda \Theta_{\rho\sigma}
	\left(\frac{1}{3}\eta^\mu_{\;\tau}+\frac{2}{3}u^\mu u_\tau\right)\frac{T^3(x)}{12}\\
= & u^\mu \epsilon^{\lambda\rho\sigma\tau}\hat{t}_\lambda\Theta_{\rho\sigma}u_\tau\frac{T^3(x)}{36}+\\
	&-\frac{1}{2}\epsilon^{\mu\rho\sigma\lambda}\Theta_{\rho\sigma}\hat{t}_\lambda \frac{T^3(x)}{9},
\end{split}
\end{equation*}
where I took advantage of the Lorentz invariance of the integral measure and I chose the unit vector $u=\beta/\sqrt{\beta^2}$ as the
time direction for the $k$ four-vector. Furthermore, if one decomposes $\hat{t}$ in a orthogonal and parallel parts respect to $u$ as
\begin{equation*}
\hat{t}_\lambda=(\hat{t}\cdot u) u_\lambda + \hat{t}_{\perp\lambda}
\end{equation*}
and uses $\Theta_{\rho\sigma}=\varpi_{\rho\sigma}-\Omega_{\rho\sigma}$ and the definition~\eqref{eq:thermalrot}, one obtains
\begin{equation*}
\begin{split}
\Delta\sups{C}_\Theta j^\mu\subs{A}(x)= & u^\mu \epsilon^{\lambda\rho\sigma\tau}\hat{t}_\lambda\Theta_{\rho\sigma}u_\tau\frac{T^3(x)}{36}+\\
	&+w^\mu\frac{(\hat{t}\cdot u)T^3(x)}{9} -\frac{1}{2}\epsilon^{\mu\rho\sigma\lambda}\Theta_{\rho\sigma}\hat{t}_{\perp\lambda} \frac{T^3(x)}{9}+\\
	&+\frac{1}{2}\epsilon^{\mu\rho\sigma\lambda}\Omega_{\rho\sigma}\hat{t}_\lambda \frac{T^3(x)}{9}.
\end{split}
\end{equation*}
The second term describes a mean axial current flowing along the rotation of the fluid and therefore gives a modification of the AVE conductivity of
\begin{equation*}
\Delta\sups{C}_\Theta W\sups{A}= \frac{(\hat{t}\cdot u)T^3(x)}{9}.
\end{equation*}
In general, the modification of the AVE conductivity for the other pseudogauges in both the massive and massless cases
can be obtained from the general expressions~\eqref{eq:AxialPGDep} by projecting along $w$:
\begin{equation*}
\Delta_\Phi W\sups{A}= \frac{w_\mu}{w^2}\left(\Delta_\Theta j^\mu\subs{A}(x) + \Delta_\xi j^\mu\subs{A}(x) \right).
\end{equation*}
%

\section{Discussion}
\label{sec:conclusion}
In summary, I showed that the pseudogauge (PG) dependent part of the statistical operator describing
a system at local thermal equilibrium (LTE) results in a non-vanishing contribution to the Wigner function
of the free Dirac field. I then evaluated the spin polarization at LTE with different choices of
pseudogauge potentials and I found that they generally give different results.
In the canonical case, the contribution from the canonical spin tensor is different
form what one might expect by just replacing the thermal vorticity with the spin potential in Eq.~\eqref{eq:PolBelIntro},
see Eq.~\eqref{eq:DeltaSCan}. The de Groot-van Leeuwen-van Weert (GLW) and the Hilgevoord-Wouthuysen (HW) decompositions
result in the same spin polarization. This spin polarization differs from the ones resulting from the Belinfante
and from the canonical decomposition. In particular, there is no contribution from thermal-shear in the GLW and HW PG,
see Eq.~\eqref{eq:DeltaXiGLW}. It is important to stress that this
difference does not depend on the spin potential. Therefore, the choice of the pseudogauge significantly affects
the spin polarization predictions used in heavy-ion collisions.
I also showed that the PG dependence in the spin polarization can not be removed by choosing
different values for the four-temperature $\beta(x)$ and for the spin potential $\Omega(x)$ in the different PGs.
On the contrary, as suggested by the analysis of the statistical operator itself, I found that the different PGs
give the same spin polarization only if the same $\beta$ field is chosen and if the conditions of global thermal
equilibrium are satisfied.

In principle, the matching of different predictions with experimental measurements should reveal what
pseudogauge must be used and if this choice is universal for all systems. However, given the
arbitrariness of pseudogauge transformations, this method alone does not seem compelling. I advocate
instead for the search of a theoretical argument in favor of a particular pseudogauge, which one can
thereafter put to test.  When these predictions are compared with experiments,
one should also keep in mind that they are the result of the approximation of the ``true'' statistical
operator~\eqref{eq:statoperTrue} with the local equilibrium one~\eqref{eq:statoperLTE} which only
account for non-dissipative phenomena. The dissipative effects for the spin polarization are
yet unknown. Future studies on that topic are therefore recommended.

More stringent requirements on the pseudogauge transformations might come from the theory of general
relativity, where the form of the energy-momentum tensor is strictly related to the geometry.
The Einstein-Cartan theory allows the inclusion of a spin tensor in general relativity and requires
the use of Reimann-Cartan geometry~\cite{Hehl:1976vr}, which has non-vanishing torsion. Then,
the comparison of results from quantum hydrodynamics, gravitation, and proton spin decomposition~\cite{Ji:2020ena}
might shed light on this long standing problem.

Furthermore, I showed that the pseudogauge transformations also affect the axial vortical effect (AVE)
conductivity for a system out of global equilibrium. This might pose a problem to the interpretation
of the AVE as a consequence of the gravitational anomaly~\cite{Landsteiner:2011cp}, as one might expect
the gravitational anomaly to be universal and not pseudogauge dependent. Further investigation in this
direction might also clarify the role of spin in gravity and hydrodynamics.
	
\acknowledgments
Very useful discussions with F.~Becattini and K.~Tuchin are gratefully acknowledged.
I am also immensely grateful to N. Weickgenannt for her comments,
that helped me identify some mistakes in the calculations.
M.B. is supported by the U.S. Department of Energy under Grant No. DE-FG02-87ER40371.
\appendix
\section{Pseudogauge operators}
\label{sec:AppA}
In this appendix I report the different form of pseudogauge fields used to perform the calculations.
For the Dirac field, the pseudogauge operators in~(\ref{eq:PGPotentialsGamma} - \ref{eq:PGPotentialHW}) can be obtained
from the Wigner operator
\begin{equation*}
\h{W}_{AB}(x,k)=\!\!\int\!\frac{\D^4y}{(2\pi)^4}\E^{-\I k\cdot y}:\overline{\Psi}_B \left( x+\frac{y}{2}\right)\Psi_A\left( x-\frac{y}{2}\right):
\end{equation*}
by performing the integration
\begin{equation*}
	\wPhi^{\lambda,\mu\nu}(x)=\int\D^4k' \Phi^{\lambda,\mu\nu}_{B'A'}(k') \h{W}_{A'B'}(x,k').
\end{equation*}
For the gaugepotentials in Eq.~(\ref{eq:PGPotentialsGamma}) the function $\Phi(k)$ is given by
\begin{equation*}
\begin{split}
\Phi^{\lambda,\rho\sigma}\subs{C}= & \Phi^{\lambda,\rho\sigma}_{\epsilon{\rm A}}=\frac{1}{2}\epsilon^{\lambda\rho\sigma\tau}\gamma_\tau \gamma^5,\\
\Phi^{\lambda,\rho\sigma}_{\epsilon{\rm V}}= & \frac{1}{2}\epsilon^{\lambda\rho\sigma\tau}\gamma_\tau,\\
\Phi^{\lambda,\rho\sigma}\subs{A}= & \left(\eta^{\lambda\rho}\gamma^\sigma-\eta^{\lambda\sigma}\gamma^\rho\right)\gamma^5,\\
\Phi^{\lambda,\rho\sigma}\subs{V}= & \eta^{\lambda\rho}\gamma^\sigma-\eta^{\lambda\sigma}\gamma^\rho,
\end{split}
\end{equation*}
for those in Eq.~\eqref{eq:PGPotentialsDe} it is
\begin{equation*}
\begin{split}
\Phi^{\lambda,\rho\sigma}_{\epsilon\de}= & 2\epsilon^{\lambda\tau\rho\sigma} k_\tau,\\
\Phi^{\lambda,\rho\sigma}_\de= & 2\left(\eta^{\lambda\rho}k^\sigma - \eta^{\lambda\sigma}k^\rho\right),\\
\Phi^{\lambda,\rho\sigma}_{\epsilon\de {\rm A}}= & 2\left(\eta^{\lambda\rho}k^\sigma - \eta^{\lambda\sigma}k^\rho\right)\gamma^5,\\
\Phi^{\lambda,\rho\sigma}_{\de {\rm A}}= & \epsilon^{\lambda\tau\rho\sigma}k_\tau\gamma^5,\\
\Phi^{\lambda,\rho\sigma}_{\de\Sigma}= & \frac{2}{m}k^\lambda\sigma^{\rho\sigma},
\end{split}
\end{equation*}
for the de Groot-van Leeuwen-van Weert (GLW) decomposition~(\ref{eq:PGPotentialGLW}) it is
\begin{equation*}
\Phi^{\lambda,\rho\sigma}\subs{GLW}=\Phi^{\lambda,\rho\sigma}_{\epsilon{\rm A}}+\frac{1}{2m}\left(\sigma^{\lambda\rho}k^\sigma-\sigma^{\lambda\sigma}k^\rho\right)
\end{equation*}
and for the Hilgevoord-Wouthuysen decomposition of Eq.~(\ref{eq:PGPotentialHW}) it is
\begin{equation*}
\Phi^{\lambda,\rho\sigma}\subs{HW}=\Phi^{\lambda,\rho\sigma}\subs{GLW}-\frac{1}{2m}\left(\eta^{\lambda\rho}\sigma^{\sigma\alpha}-\eta^{\lambda\sigma}\sigma^{\rho\alpha}\right)k_\alpha .
\end{equation*}
The improved Callan, Coleman, and Jackiew (CCJ) pseudogauge potentials~\eqref{eq:PGPotentialCCJ} are reported in Eq.~\eqref{eq:CCJPotentials}.

In order to obtain the difference between the Wigner function in a generic pseudogauge and the Wigner function in the Belinfante pseudogauge,
one just has to replace the right expression above in Eq.~\eqref{eq:DeltaWigPseudoFinal}. 
To obtain the corrections to spin polarization~\eqref{eq:DeltaS} and to the mean axial current~\eqref{eq:AxialPGDep},
the expression above must be replaced inside the traces,
\begin{equation*}
\begin{split}
A^{\mu\rho\sigma\lambda}_{\Theta,\,\Phi}=&\tr_4\left[\gamma^\mu\gamma^5(\slashed{k}+m)\Phi^{\lambda,\rho\sigma}(\slashed{k}+m) \right], \\
A^{\mu\rho\sigma\lambda}_{\xi,\,\Phi}=&\tr_4\left[\gamma^\mu\gamma^5(\slashed{k}+m)\Phi^{\rho,\lambda\sigma}(\slashed{k}+m) \right],
\end{split}
\end{equation*}
whose results are given in Eqs.~\eqref{eq:ATheta} and \eqref{eq:AXi}.
\end{document}